\documentclass[%
 reprint,
superscriptaddress,
 amsmath,amssymb,
pra,
floatfix,
]{revtex4-2}

\usepackage{graphicx}
\usepackage{dcolumn}
\usepackage{bm}
\usepackage{color}
\usepackage{upgreek}
\usepackage{graphicx}
\usepackage{subcaption}
\usepackage{ragged2e}       
\usepackage{caption}        
\usepackage[utf8]{inputenc}
\definecolor{rc}{RGB}{1,0,0}

\begin{document}

\preprint{APS/123-QED}

\title{Interplay of Flat-band and Anderson localizations in disordered moiré superlattices}

\author{Qian Liu}
\affiliation{%
    Guangxi Key Lab for Relativistic Astrophysics, Center on Nanoenergy Research, School of Physical Science and Technology, Guangxi University, Nanning, Guangxi 530004, China}%
\author{Xiaoshuang Xia}%
\affiliation{%
     Guangxi Key Lab for Relativistic Astrophysics, Center on Nanoenergy Research, School of Physical Science and Technology, Guangxi University, Nanning, Guangxi 530004, China}%
\author{Junjie Wang}
\affiliation{%
    Guangxi Key Lab for Relativistic Astrophysics, Center on Nanoenergy Research, School of Physical Science and Technology, Guangxi University, Nanning, Guangxi 530004, China}%
\author{Peilong Hong}
\email{p.l.hong@aqnu.edu.cn}
\affiliation{%
    School of Mathematics and Physics, Anqing Normal University, Anqing, 246133 Anhui, China}%
\affiliation{%
    The MOE Key Laboratory of Weak-Light Nonlinear Photonics, School of Physics and TEDA Applied Physics Institute, Nankai University, Tianjin 300457, China}%
    
\author{Lei Xu}
\email{ lei.xu@ntu.ac.uk}
\affiliation{%
    Advanced Optics \& Photonics Laboratory, Department of Engineering, School of Science \& Technology, Nottingham Trent University, Nottingham NG11 8NS, UK}%
    
\author{Lujun Huang}
\affiliation{%
    School of Physics and Electronic Science, East China Normal University, Shanghai 200241, China}%
    
\author{Daohong Song}
\affiliation{%
    The MOE Key Laboratory of Weak-Light Nonlinear Photonics, School of Physics and TEDA Applied Physics Institute, Nankai University, Tianjin 300457, China}%
\author{Yi Liang}
\email{liangyi@gxu.edu.cn}
\affiliation{%
    Guangxi Key Lab for Relativistic Astrophysics, Center on Nanoenergy Research, School of Physical Science and Technology, Guangxi University, Nanning, Guangxi 530004, China}%

\date{\today}

\begin{abstract}
Disorder in moiré superlattices simultaneously degrades flat‑band localization and induces Anderson localization, yet how these two regimes interact has remained unclear. Here, we introduce a combined framework linking localization‐length scaling with differential probability density analysis to map localization transitions in partially disordered one‐dimensional silicon moiré lattices. It is found that flat bands confined within the interband gap keep their strong localization even as disorder grows. In contrast, flat bands intersecting dispersive bands exhibit rich behaviors: the low‐frequency branch undergoes an inverse Anderson transition, while the high‐frequency branch supports coexisting flat‑band and Anderson localization at strong disorder. Our results deliver the direct evidence of competing localization mechanisms in disordered moiré systems and offer guiding principles for engineering robust, nonideal moiré photonic devices.

\end{abstract}

\maketitle

 Moiré photonic crystals have emerged as a powerful platform for engineering light. They host flat bands at magic twist angle~\cite{dong2021flat}, and realize light localization~\cite{wang2020localization} without defects~\cite{makasyuk2006band}, disorder~\cite{rosen2024flat}, or nonlinearity~\cite{fleischer2003observation}, as a result of coherent phase cancellation among Bloch wavefunctions and the macroscopic degeneracy of eigenmodes~\cite{shi2020novel}. Moiré photonic flat bands enable unprecedented enhancement of light-matter interactions by achieving ultra-high photon density of states and robust spatial localization, revolutionizing the design of high-efficiency quantum light sources and compact photonic devices~\cite{Luan2023,Yu2025,hong2024high}. 

Flat bands are typically categorized into two classes with distinct topological and localization characteristics ~[\onlinecite{rhim2019classification}]: non‑singular flat bands that lie entirely within a bandgap and singular flat bands that intersect dispersive bands at isolated points in momentum space ~[\onlinecite{bergman2008band}]. In moiré photonic crystals, both types of flat bands and their localizations can emerge and be tuned by adjusting twist angle ~[\onlinecite{mao2021magic}], interlayer spacing ~[\onlinecite{nguyen2022magic}] and band offset ~[\onlinecite{hong2023robust}]. However, practical photonic devices cannot avoid structural disorder [\onlinecite{emoto2022wide}], which tends to blur these flat‑band localizations. Conversely, strong disorder itself can induce Anderson localization, which traps light via multiple‑scattering interference ~\cite{lee2018anderson,bertolotti2005optical,segev2013anderson}. Hence, in practical moiré superlattices, flat‑band localization ~[\onlinecite{aoki1996hofstadter}] and Anderson localization ~[\onlinecite{Y2000Fundamentals}] coexist and compete. Understanding the interaction between these two localizations is crucial for controlling light confinement to design advanced micro and nanophotonic devices
~\cite{oudich2024engineered,zhang2023non,zeng2024transition,li2022aharonov}.

Here, we systematically investigate the interplay of flat-band and Anderson localizations by introducing controlled structural disorder into a moiré superlattice through randomly varying the silicon strip widths. Our results show that non-flat-band modes present a monotonic decrease in localization length as disorder grows, resulting in a transition from extended to localized states near the bandgap edges. In other words, conventional  Anderson localization appears, as demonstrated in the statistical analysis of the probability density of optical intensity. In contrast, flat-band modes exhibit markedly different behavior depending on their spectral position. Isolated flat bands residing within the coupling bandgap maintain robust spatial localization even under strong disorder. However, intersecting flat bands, which cross dispersive bands, undergo more complex dynamics: at low disorder, they exhibit an inverse Anderson transition, where increasing disorder leads to delocalization; yet at sufficiently high disorder, high-frequency intersecting flat bands re-establish localization. These unconventional localization phenomena reveal the rich physics of disordered moiré superlattices and pave the way for designing photonic devices with controlled localization.

\begin{figure}[tbp!]
\centering
\includegraphics[height=4.7cm,width=8.5cm]{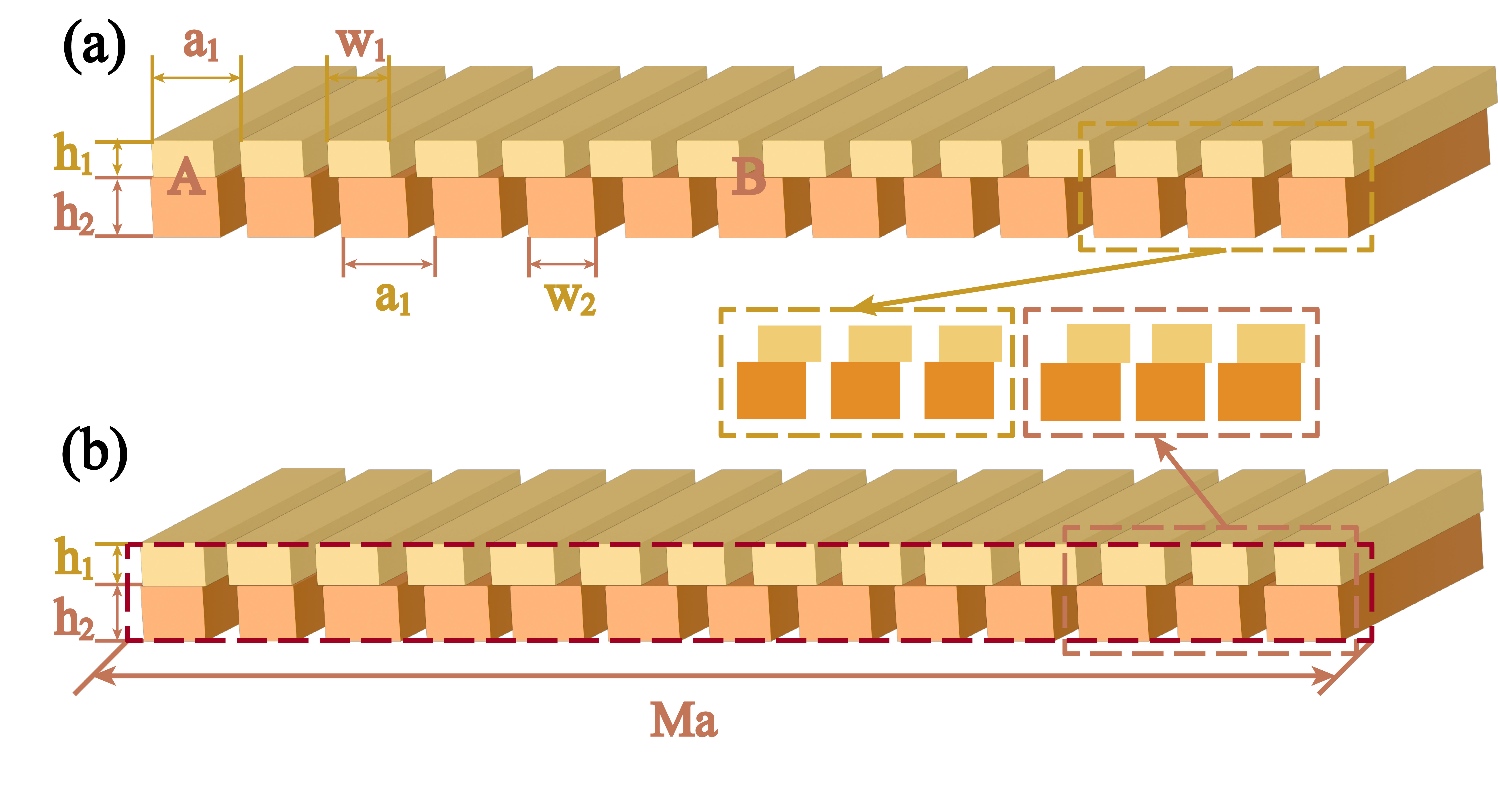}
\captionsetup{
  justification=RaggedRight, 
  singlelinecheck=false,     
  hypcap=true                
}
\caption{\label{fig:A} (a) Ideal and (b) disorder moiré superlattices.}
  \phantomsubcaption\label{fig:A:a} 
  \phantomsubcaption\label{fig:A:b} 

\end{figure}

\begin{figure}[t!]
\centering
\includegraphics[height=6.4cm,width=8.5cm]{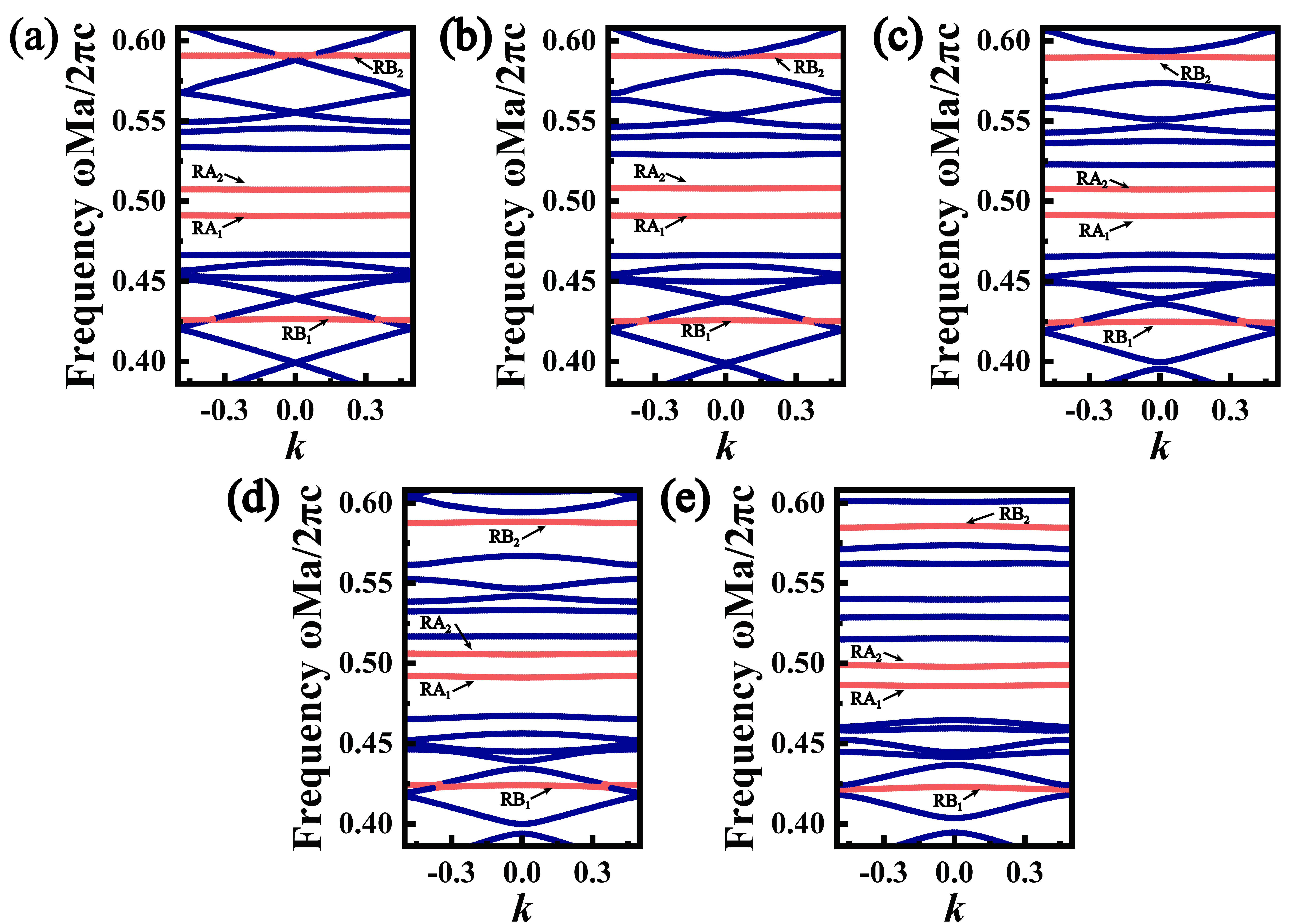}
\captionsetup{
  justification=RaggedRight, 
  singlelinecheck=false,     
  hypcap=true                
}
\caption{\label{fig:B}Band structures under varying disorder strengths. (a-e) $\alpha$ = 0 (ideal), 0.25, 0.5, 0.75, and 1, respectively. Calculated band structures using the supercell method with periodic boundary conditions.}

 \textbf{ 
  \phantomsubcaption\label{fig:B:a} 
  \phantomsubcaption\label{fig:B:b} 
  \phantomsubcaption\label{fig:B:c} 
  \phantomsubcaption\label{fig:B:d} 
  \phantomsubcaption\label{fig:B:e} 
}

\end{figure}

As shown in Fig.~\ref{fig:A}(\subref{fig:A:a}),  our moiré superlattice is constructed by superposition of two periodic silicon slabs with different spatial periods. To form a moiré superlattice, the period of slab 1 is set to $a_1 = 2N/(2N + 1)a_0$ while that of slab 2 is $a_2 = 2(N + 1)/(2N + 1)a_0$. Thus, the superlattice period is $Ma = (N + 1)a_1 = Na_2 = 2N(N+1)/(2N + 1)a_0$, where $a_0$ is $a$ constant parameter ($a_0 = 300 nm$) and $N$ is a fixed integer ($N = 13$ hereafter). Furthermore, the width of the silicon strip is configured as $w_i = 0.7a_i$ ($i = 1, 2$). The thickness of slab 1 is maintained at $h_1 = 0.4a_0$, while that of slab 2 is given by $h_2 = 0.7a_0$. Then, we calculated the energy bands for this superlattice with these parameters.  As shown in Fig.~\ref{fig:B}(\subref{fig:B:a}), two sets of flat bands (designated as $RA$ and $RB$ groups) appear. $RA$ flat bands are situated within the dispersive band gap, maintaining isolation and flatness without intersections with other dispersive bands. In contrast, $RB$ flat bands appear outside the band gap region, exhibiting crossing points with the dispersive bands.

In the following, we will study the disorder effect on the localization features of light in the superlattice. To introduce structural disorder, an independent random change is added to the width for each silicon strip, given by
\begin{eqnarray}
\Delta L_{ij} = \alpha \cdot x_{ij} \cdot L_{\text{max} },
\end{eqnarray}
where $\Delta L_{ij}$ denotes the amount of random change in the width of the $jth$ strip of the $ith$ layer, where $\alpha$ denotes the disorder degree, and $x_{ij}$ is a random number uniformly distributed between $- 1$ and $1$. $L_{max} = a_{1}-w_{2}  $, which is limited for avoiding the overlap of nearby strips after random perturbation at the maximum disorder degree ($\alpha =1$). Hence, the disorder degree for the moiré superlattice is controlled by the parameter $\alpha$, which belongs to $\left [ 0,1 \right ]$. Figure~\ref{fig:A}(\subref{fig:A:b}) shows a typical schematic diagram for a superlattice with disorder degree $\alpha =0.5$. By introducing different structure disorders, there is a significant change of the moiré bands in the spectral domain including the two sets of flat bands, as shown in Figs.~\ref{fig:B}(\subref{fig:B:b}-\subref{fig:B:d}). By employing ensembles of disordered moiré superlattice at different disorder degrees, we next investigate the interplay of Anderson localization and flat-band localization in the spatial domain.

 The optical response of any medium can be described by a set of eigenmodes $\psi_{n} $ with eigenfrequencies $\omega _{n} $  through solving Maxwell's equations with appropriate boundary conditions. To illustrate the effect of disorder on the optical localization in moiré photonic superlattices, we have calculated the eigenmodes via full-wave simulations ~\cite{escalante2018level,lourtioz2005photonic,joannopoulos2008molding}, which is expressed as
\begin{eqnarray}
\psi _{n,\mathbf{k} }\left ( \mathbf{r}   \right )  =  u_{n,\mathbf{k} } (\mathbf{r} )exp(i\mathbf{k\cdot r } ),
\end{eqnarray}
where $u_{n,\mathbf{k}} (\mathbf{r} )$ denotes the Bloch wavefunction at wave vector $\mathbf{k }$, and $\mathbf{r }$ represents the spatial coordinate.
Then, we can characterize their localization via calculating the well-established inverse participation ratio (IPR) ~[\onlinecite{escalante2018level}]:
\begin{eqnarray}
IPR=\frac{\int_{A}^{}\left | u_{n,\mathbf{k} } (\mathbf{r} ) \right |^{4} d^{2} \mathbf{r}}{\left [ \int_{A}^{}\left | u_{n,\mathbf{k} } (\mathbf{r} ) \right |^{2} d^{2} \mathbf{r} \right ] ^{2} } 
\end{eqnarray}
and  the localization length of each eigenmode:
\begin{eqnarray}
\xi _{n,\mathbf{k}}=\frac{1}{2}\sqrt{\frac{1}{IPR} } .
\end{eqnarray}
The localization length characterizes the spatial extent of the eigenmode profile, where a smaller $\xi _{n,\mathbf{k}}$ indicates stronger localization. In the following analysis, we compare the $\xi _{n,\mathbf{k}}$ values for systems with varying disorder.

\begin{figure}[t!]
\includegraphics[height=6.5cm,width=8.5cm]{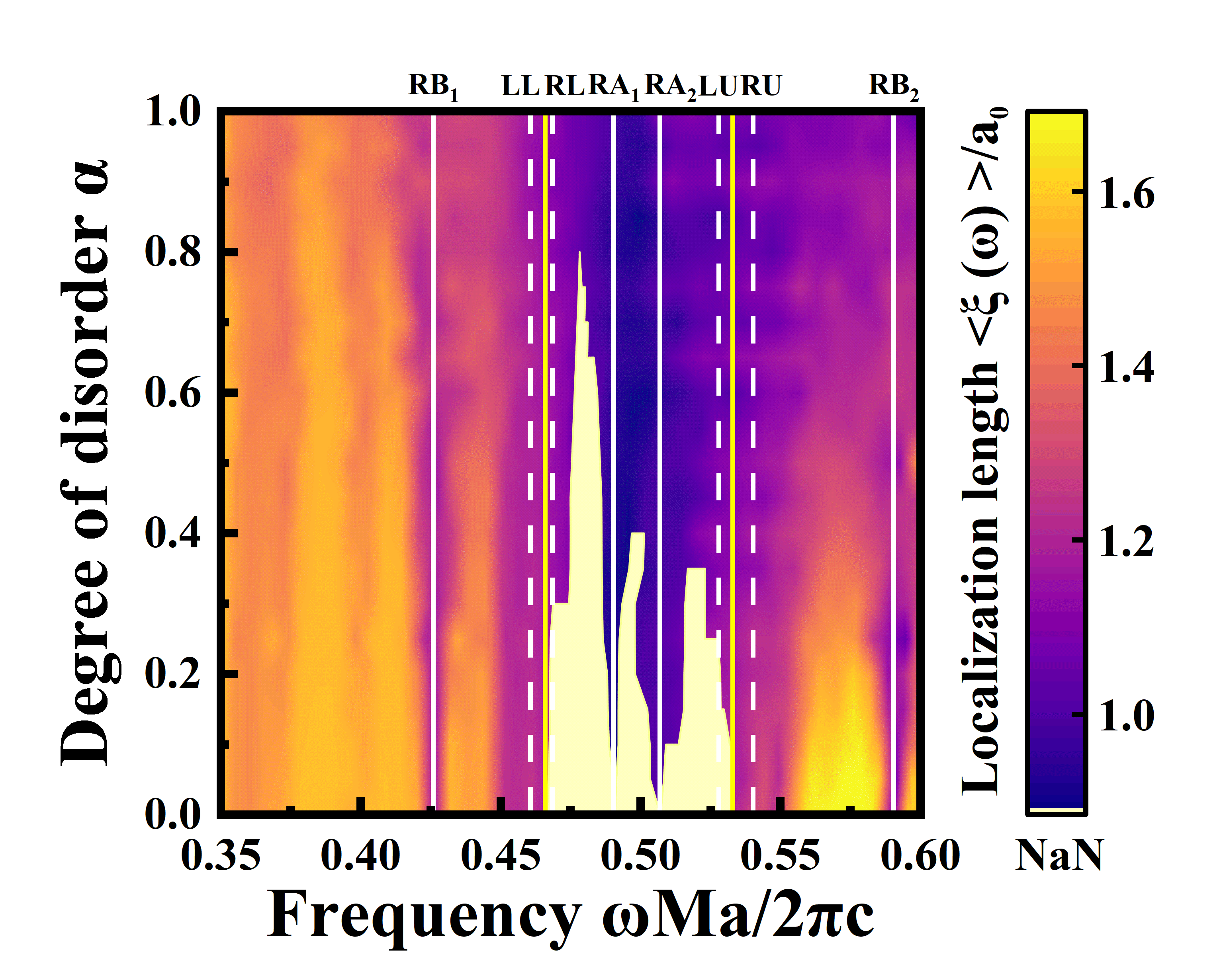}
\captionsetup{
  justification=RaggedRight, 
  singlelinecheck=false,     
  hypcap=true                
}
\caption{\label{fig:C}Average localization lengths for TE modes at different disorder degrees $\alpha = 0\sim 1$. Yellow solid lines mark the band gaps of the non-flat-bands at $\alpha = 0$, dividing the spectrum into regions I-III. White dashed lines show the positions of the non-flat-bands near the band gap edges. White solid lines indicate the two sets of flat-bands, respectively. Pale yellow areas present the band gaps with no localization.}
\end{figure}

In disordered systems, $\xi _{n,\mathbf{k}}$ varies across different disorder configurations. To quantify the localization length physically, we compute the ensemble average $\left \langle \xi\left ( \omega  \right )  \right \rangle $ ~[\onlinecite{escalante2018level}]. For each disorder degree, we simulate 100 distinct configurations of disordered superlattices, focusing on TE modes containing two flat-band sets ~[\onlinecite{hong2023robust}]. For sufficient averaging, the averaged $\left \langle \xi\left ( \omega  \right )  \right \rangle $ is evaluated within a frequency interval of width $\bigtriangleup \omega Ma/2\pi c= 0.006$ 
 centered at $\omega $. Figure~\ref{fig:C} displays the normalized average localization length $\left \langle \xi \left ( \omega  \right )  \right \rangle /a_{0} $ for TE modes under different disorder degrees, with the spectrum partitioned into three regions: below (I), within (II), and above (III) the band gap of an ideal superlattice. In addition to marking the positions of the four flat bands in the figure, we have also labeled the positions of four flat bands, specifically on the left side of the lower bandgap($LL$), the right side of the lower bandgap($RL$), the left side of the upper bandgap($LU$), and the right side of the upper bandgap($RU$).

As shown in Fig.~\ref{fig:C}, when the disorder degree $\alpha$ increases, the gap (region II) becomes progressively narrower and eventually closes approaching $\alpha = 1$. It is a result of structure disorder that stronger perturbation drags optical modes from the upper and lower energy bands into band gap. For non-flat-band positions near the band edges, the $\left \langle \xi\left ( \omega  \right )  \right \rangle $ exhibits a typical Anderson-type localization, i.e. stronger disorder (larger $\alpha$) yields shorter $\left \langle \xi\left ( \omega  \right )  \right \rangle $. In contrast, the $\left \langle \xi\left ( \omega  \right )  \right \rangle $  at flat-band positions exhibits complex trends:

i) As disorder degree increases, the color at positions $RB_{1}$ and $RA_{1}$ gradually becomes lighter, indicating an increase in $\left \langle \xi\left ( \omega  \right )  \right \rangle $ and reduced localization. This demonstrates a typical inverse Anderson trend.

ii) At position $RA_{2}$, the color initially deepens under low disorder but subsequently becomes lighter as disorder increases, implying a non-monotonic variation of $\left \langle \xi\left ( \omega  \right )  \right \rangle $ at this site.

iii) For position $RB_{2}$, the color initially becomes lighter, with an increase in $\left \langle \xi\left ( \omega  \right )  \right \rangle $, showing an inverse Anderson trend. Subsequently, as disorder increases further, the color progressively intensifies, $\left \langle \xi\left ( \omega  \right )  \right \rangle $ decreases, and the trend transits to Anderson localization.

\begin{figure}[htbp]
\includegraphics[height=5.6cm,width=8.5cm]{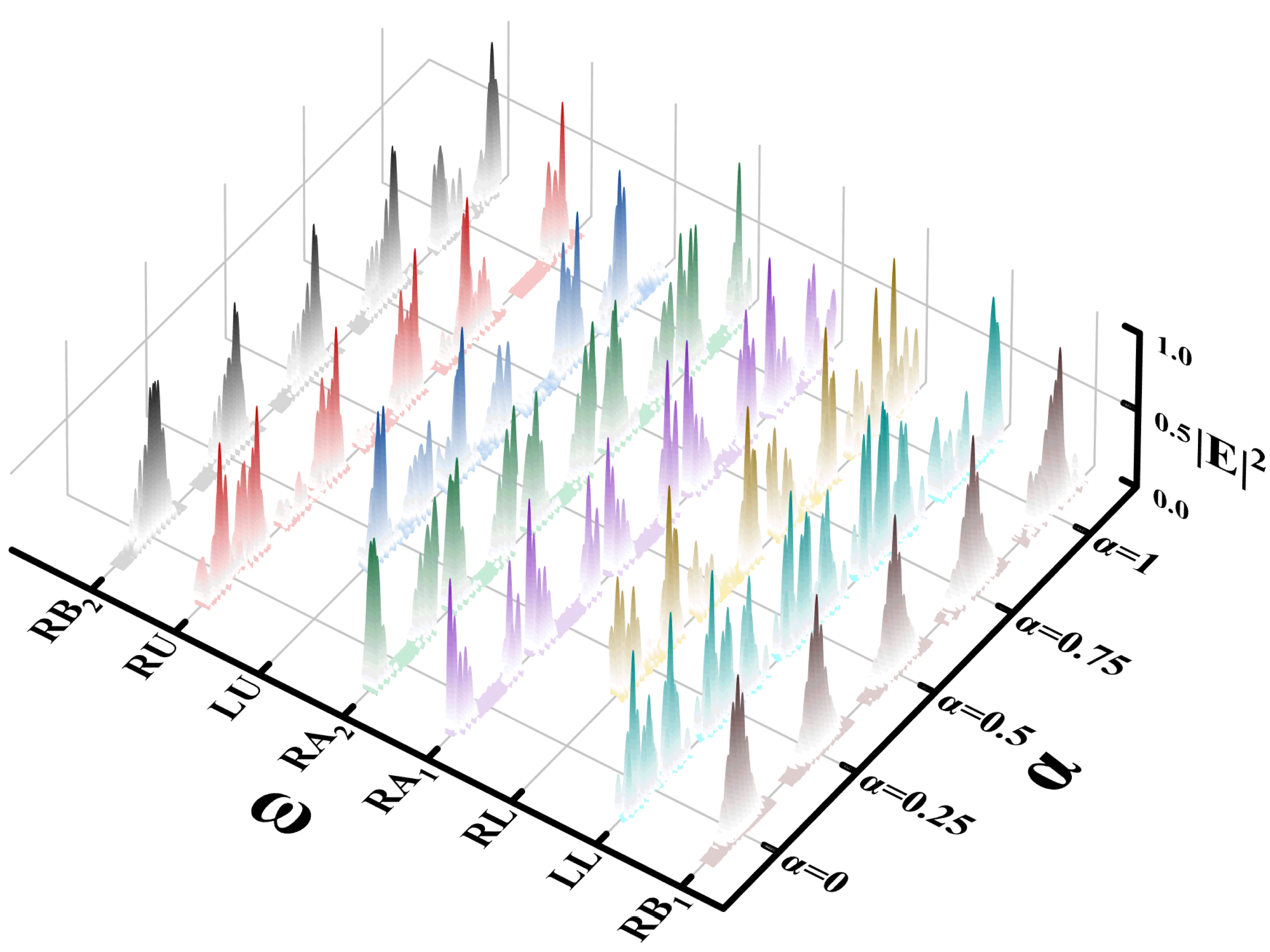}
\captionsetup{
  justification=RaggedRight, 
  singlelinecheck=false,     
  hypcap=true                
}
\caption{\label{fig:D}Typical intensity distributions of TE modes for different disorders and frequencies at $\mathit{k}  =0.3$. At zero disorder, $RL$ and $LU$ modes do not appear. As disorder increases, $RA$ flat-band modes stay localized, $RB_{1}$ becomes more extended, and $RB_{2}$ changes from localized to extended, then back to localized at high disorder. Other non-flat-band modes gradually become localized as disorder grows.}

\end{figure}

Figure~\ref{fig:D} presents the intensity distributions for TE modes at $\mathit{k}  =0.3$ under varying disorder degrees and frequencies, in agreement with the $\left \langle \xi\left ( \omega  \right )  \right \rangle $ results. Although $\left \langle \xi\left ( \omega  \right )  \right \rangle $ captures relative changes in spatial localization, it can not identify the critical disorder where localization transits to delocalization . For example, if the system is partitioned into $M$ regions of size $\xi /\sqrt{M} $ separated by arbitrarily large distances, $\left \langle \xi\left ( \omega  \right )  \right \rangle $ remains unchanged ~[\onlinecite{escalante2018level}], underscoring its limitations in fully characterizing spatial localization. Therefore, $\left \langle \xi\left ( \omega  \right )  \right \rangle $ alone cannot serve as a unique metric for localized states. 

To address this limitation and gain deeper insights into the interplay between flat-band and Anderson localizations, we employ the probability density distribution of intensity for identifying the transition between localization and delocalization, providing a more robust framework for characterizing the localization and delocalization of scattered modes in partially disordered moiré superlattices.

 The intensity probability $P\left ( \frac{I}{\left \langle I \right \rangle }  \right ) $  serves as a key metric for characterizing complex light fields. For random light fields, Rayleigh speckle patterns represent the canonical example of scattered light. In Rayleigh-type random fields, the optical intensity follows an exponential probability density distribution ~[\onlinecite{han2023tailoring}]
\begin{eqnarray}
P\left ( \frac{I}{\left \langle I \right \rangle }  \right ) =exp\left ( \frac{-I}{\left \langle I \right \rangle }  \right ).
\end{eqnarray}

Given that Rayleigh-type fields arise from light interference along random scattering paths ~[\onlinecite{goodman2007speckle}], we propose Eq. (5) as a robust criterion to distinguish between localized and delocalized states. For localized states, light is concentrated at specific spatial points while remaining dark elsewhere, resulting in suppressed probabilities at intermediate intensities and producing a characteristic dip in the probability distribution. Conversely, delocalized states exhibit dominant intermediate intensities, forming a single peak of probability. 

\begin{figure}[b!]
\includegraphics[height=7.2cm,width=8.5cm]{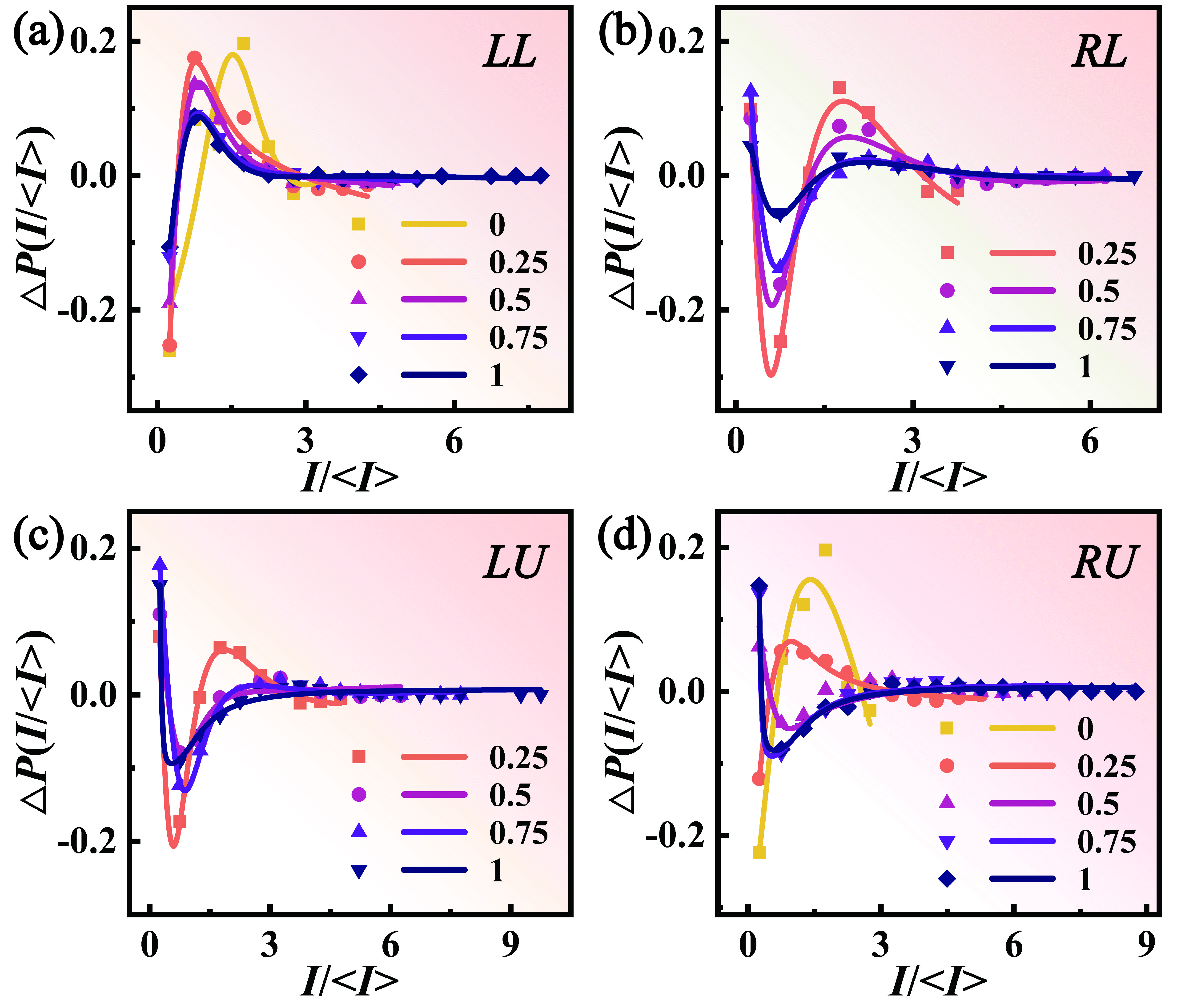}
\captionsetup{
  justification=RaggedRight, 
  singlelinecheck=false,     
  hypcap=true                
}
\caption{\label{fig:E}Statistical Differential probability density $\bigtriangleup P$ for TE modes corresponding to $LL$,$RL$,$LU$, and $RU$ under different disorder degrees $\alpha = 0\text{-}1$. (a, b)  show the localization of the lower edge of the band gap (left and right non-flat bands) gradually strengthens with increasing disorder. (c, d) show the upper edge (non-flat bands on both sides). In (d), strong localization appears at $\alpha = 0.5$, indicating Anderson localization. Since disorder introduces new modes inside the band gap, there are no data for $\alpha = 0$ within the gap region, as seen in (b, c).}

  \phantomsubcaption\label{fig:E:a} 
  \phantomsubcaption\label{fig:E:b} 
  \phantomsubcaption\label{fig:E:c} 
  \phantomsubcaption\label{fig:E:d} 

\end{figure}

To compute $P\left ( \frac{I}{\left \langle I \right \rangle }  \right ) $, we simulated 100 disorder configurations at a fixed disorder degree $\alpha $ at $\mathit{k}  =0.3 $. To quantify spatial localization along the x-axis, the y-axis intensity distributions were averaged for each configuration. The intensities were normalized by the ensemble-averaged intensity $\left \langle I \right \rangle $, yielding the dimensionless quantity $\frac{I}{\left \langle I \right \rangle } $.
Subsequently, the intensity probability density at each position is calculated. Finally, we obtained the differential probability density $\bigtriangleup P$ by subtract the value of the Rayleigh probability density ($P_{Rayleigh} $ ).
\begin{eqnarray}
\bigtriangleup P= P_{Measured}-P_{Rayleigh} .
\end{eqnarray}

We first examine spectral regions near band gap edges: $LL$, $RL$, $LU$, and $RU$. Since the distinction between localized and delocalized states remains invariant under variations of the wave vector $\mathbf{k }$, we present the differential probability density $\bigtriangleup P\left ( \frac{I}{\left \langle I \right \rangle }  \right ) $ at $\mathit{k}  =0.3 $ for these non-flat bands. Analysis of disorder-induced modes shows that there is no Anderson localization at $LL$ position with increasing $\alpha $ [Fig.~\ref{fig:E}(\subref{fig:E:a})]. This behavior stems from the relative insensitivity of low-frequency modes to disorder perturbations, where eigenstates remain weakly affected by broken lattice periodicity. For the two positions ($RL$ and $LU$) within the bandgap , the intensity probability density gradually transits to a dip  (signature of localized states) [Figs.~\ref{fig:E}(\subref{fig:E:b},\subref{fig:E:c})], indicating a dynamic extended-to-localized phase transition. In contrast, the upper band gap edge position ($RU$) exhibits a regular diffusive-to-localized transition [Fig.~\ref{fig:E}(\subref{fig:E:d})]. These results clearly demonstrates the emergence of conventional Anderson localization near band gaps, in agreement with previous results in 2D photonic crystals ~[\onlinecite{wiersma1997localization}].

\begin{figure}[htbp]
\includegraphics[height=7.2cm,width=8.5cm]{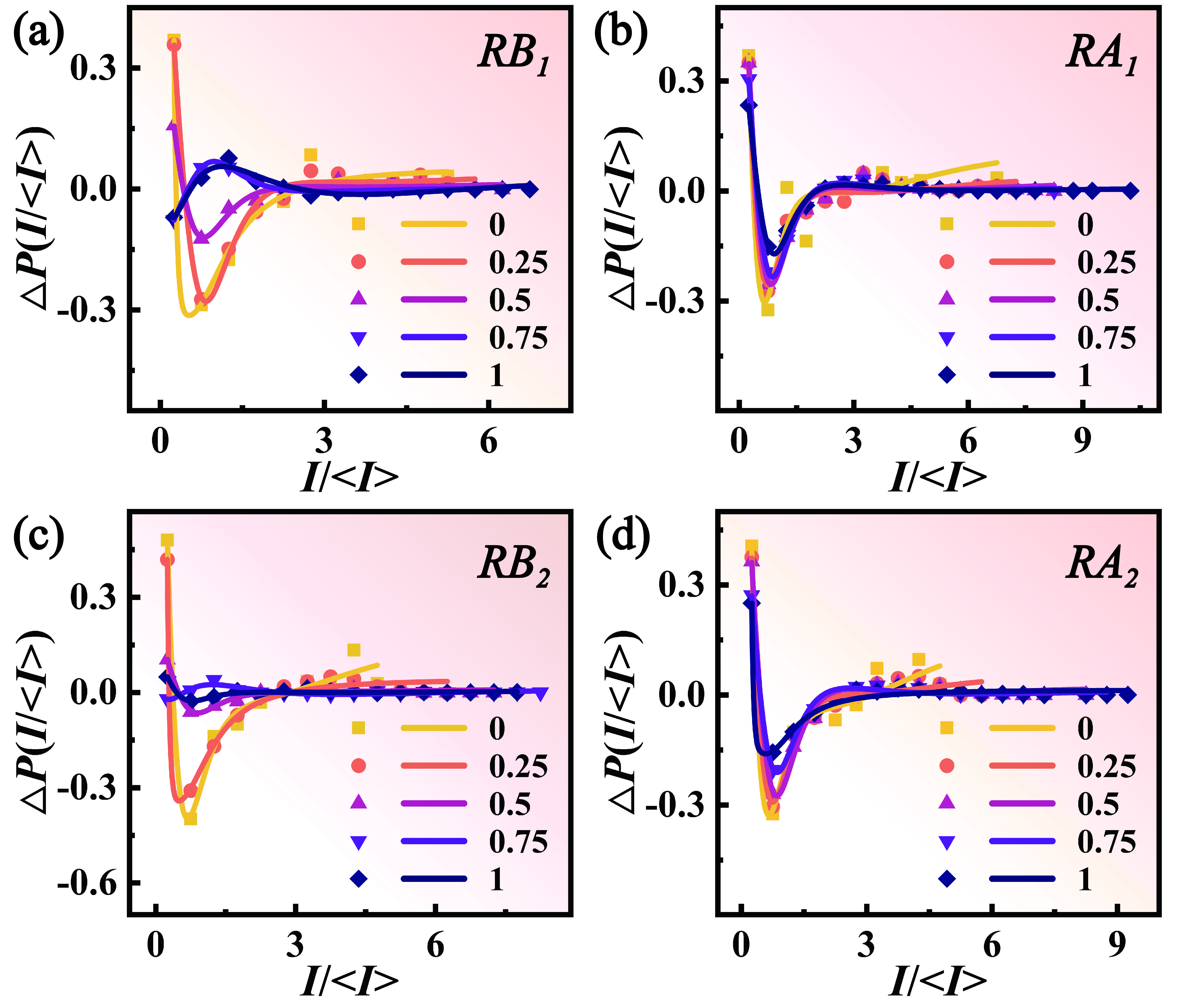}
\captionsetup{
  justification=RaggedRight, 
  singlelinecheck=false,     
  hypcap=true                
}
\caption{\label{fig:F}Statistical differential probability density $\bigtriangleup P$ for different flatbands under increasing disorder. For the $RB$ group, disorder causes a transition from localization to delocalization (inverse Anderson transition). The $RB_{2}$ band becomes localized again at high disorder. In contrast, the $RA$ group shows reduced localization but still retains localized behavior even under strong disorder.}

  \phantomsubcaption\label{fig:F:a} 
  \phantomsubcaption\label{fig:F:b} 
  \phantomsubcaption\label{fig:F:c} 
  \phantomsubcaption\label{fig:F:d} 

\end{figure}

Next, we move to the four flat bands: $RA_{1}$, $RA_{2}$, $RB_{1}$, and $RB_{2}$. Figure~\ref{fig:F} reveals a distinct disorder-correlated trend between $RA$ and $RB$ flat bands. For isolated band-gap flatbands ($RA_{1, 2}$, Figs.~\ref{fig:F}(\subref{fig:F:b}, \subref{fig:F:d})), they exhibit a typical dip of differential probability density, although the dip depth progressively decreases with increasing disorder. It means flat-band localization is maintained, but localization length becomes larger.  For dispersive-band-intersecting $RB_{1}$ flat band [Fig.~\ref{fig:F}(\subref{fig:E:a})], a clear transition of dip to peak emerges between $\alpha = 0.5$ and $\alpha = 0.75$, indicating that flat-band localization is totally destroyed at sufficiently large disorder. Remarkably, $RB_{2}$ flat band [Fig.~\ref{fig:F}(\subref{fig:F:c})] not only exhibit transition of dip to peak between $\alpha = 0.5$ and $\alpha = 0.75$ but also displays reentrant localization between $\alpha = 0.75$ and $\alpha = 1$. This return to localization implies that the interplay between flat-band and Anderson localizations could effectively suppress extended waves in disordered moiré superlattices. Thus, a significant difference in localization behavior merges in the wave localization for $RB_{1}$ and $RB_{2}$ flat bands. 

Simultaneously, we also calculated the probability density at $\mathit{k}  =0.07$ (see Supplementary Material), confirming that variations in dispersion-band distributions near flat bands do not alter the aforementioned phenomena.

In summary, we have explored the critical role of disorder in wave localization within moiré superlattices. To overcome the shortage of the well-established analysis of average localization lengths, we develop a new analytical framework based on random scattering and probability density distributions, which could precisely characterize the transition between localization and delocalization. Our results clearly demonstrate divergent localization behaviors between flat-band and non-flat-band modes near band gap edges under varying disorder degrees, including the disorder-driven Anderson localization in non-flat-band regions, persistence of flat-band localization for the $RA$-group flat bands, the localization-to-delocalization of the $RB_{1}$ flat band, and the recurrent of localization of the $RB_{2}$ flat band. These findings uncover unique localization and delocalization phenomena in disordered moiré systems, providing critical insights for designing photonic devices with engineered localization properties.

\bigskip
\textit{Acknowledgments---}This work was supported by Guangxi Natural Science Foundation (2024GXNSFAA0103144), National Natural Science Foundation of China (12474290), Innovation Project of Guangxi Graduate Education (YCSW2025064), Sichuan Science and Technology Program (2023NSFSC0460), Open Project Funding of the Ministry of Education Key Laboratory of Weak-Light Nonlinear Photonics (OS22-1) and the special funding for Guangxi Bagui Youth Scholars (Yi Liang).

\nocite{*}

\bibliographystyle{apsrev4-2}
\bibliography{apssamp}

\end{document}